\documentclass[%
reprint,
superscriptaddress,
amsmath,amssymb,
aps,
pra,
]{revtex4-2}

\usepackage{graphicx}
\usepackage{subfig}
\usepackage{dcolumn}
\usepackage{bm}

\usepackage[dvipsnames]{xcolor}
\colorlet{nblue}{blue!75!cyan}
\colorlet{nred}{magenta!90!red}
\usepackage[colorlinks,linkcolor=JungleGreen,citecolor=nblue,urlcolor=JungleGreen]{hyperref}

\newcommand{\beq}{\begin{equation}}
\newcommand{\eeq}{\end{equation}}

\newcommand{\vect}[1]{\mathbf{#1}}
\newcommand{\Eq}[1]{Eq.~\eqref{#1}}

\newcommand{\I}{\mathrm{i}}

\newcommand{\D}{\mathrm{d}}
\newcommand{\nn}{\nonumber}

\begin{document}	
	
\title{Scattering Hypervolume of Fermions in Two Dimensions}
\author{Zipeng Wang}
\author{Shina Tan}%
\email{shinatan@pku.edu.cn}
\affiliation{
	International Center for Quantum Materials, Peking University, Beijing 100871, China 
}
\date{\today}
	
\begin{abstract}
	We define the three-body scattering hypervolume $D_F$ for identical spin-polarized fermions in two dimensions, by considering the wave function of three such fermions colliding at zero energy and zero orbital angular momentum.
	We derive the asymptotic expansions of such a wave function when three fermions are far apart or one pair and the third fermion are far apart, and $D_F$ appears in the coefficients of such expansions.
	For weak interaction potentials, we derive an approximate formula of $D_F$ by using the Born expansion.
	We then study the shift of energy of three such fermions in a large periodic area due to $D_F$. This shift is proportional to $D_F$ times the square of the area of the triangle formed by the momenta of the fermions. We also calculate the shifts of energy and of pressure of spin-polarized two-dimensional Fermi gases due to a nonzero $D_F$ and the three-body recombination rate of spin-polarized ultracold atomic Fermi gases in two dimensions.
	
\end{abstract}
\maketitle

\section{Introduction}\label{sec:level1}
In quantum mechanics, the behavior of particles colliding with low energy depends sensitively on the dimension of space. For the zero energy $s$-wave collision of two particles in $d$-dimensional space, the wave function takes the following form:
$$
\renewcommand{\arraystretch}{1.5}\phi_{d\text{D}}=\left\{\begin{array}{ll}1-a_{d\mathrm{D}}/s^{d-2},&d>2,\\
	\ln(s/a_{2\text{D}}),&d=2,\end{array}\right.
$$
at $s>r_e$,
where $s$ is the distance between the two particles, $r_e$ is the range of the interaction potential, and $a_{d\mathrm{D}}$ is the $s$-wave scattering ``length'' in $d$-dimensional space.
For identical spin-polarized fermions, there are no two-body $s$-wave collisions. The two-body wave function for $p$-wave collisions also depends on the dimension of space.
The two-body wave function for the zero-energy $p$-wave collision in two dimensions is 
\begin{equation}\label{phi2}
	\phi^{(1\pm)}(\vect{s})=\left(\frac{s}{2}-\frac{2a_p}{\pi s}\right)e^{\pm i \theta}
\end{equation}
at $s>r_e$, where $\vect s$ is the spatial vector extending from one fermion to the other, and $\theta$ is the angle from the $+x$ direction to the direction of $\vect s$, such that $s_x=s\cos\theta$ and $s_y=s\sin\theta$.
$a_p$ is the $p$-wave scattering ``length" in 2D although its dimension is length squared.

The two-dimensional (2D) atomic Bose gases\cite{PhysRevLett.87.130402,Burger_2002,PhysRevLett.92.173003,PhysRevLett.93.180403,PhysRevLett.95.190403,Hadzibabic2006,PhysRevLett.98.080404,PhysRevLett.99.040402,PhysRevLett.102.170401,PhysRevA.82.013609,hung2011observation,Desbuquois2012} 
and the 2D atomic Fermi gases\cite{PhysRevLett.95.230401, PhysRevLett.105.030404,PhysRevLett.106.105301,PhysRevLett.106.105304,feld2011observation,PhysRevLett.108.045302,PhysRevLett.108.235302,PhysRevA.85.061604,Koschorreck2013,PhysRevLett.120.060402} have been successfully realised in experiments.
The two-dimensional Fermi gases have novel features not encountered in three dimensions.
For three identical spin-polarized fermions in 2D with a short-range interaction fine-tuned to a $p$-wave resonance, there are super Efimov bound states with orbital angular momentum quantum number $L=1$ \cite{superefimov,Volosniev_2014,Gridnev2014,gaochao2015}, and their binding energies obey a universal doubly exponential scaling \cite{superefimov}.

For three-body collisions, there are also big differences due to different spatial dimensions or different quantum statistics of the particles. The \emph{scattering hypervolume} was defined for identical bosons \cite{tan2008three}, distinguishable particles \cite{mestrom2021pwave,wang2021threebody,mestrom2021spin1}, and spin-polarized fermions \cite{wang2021fermion3D} in three dimensions. The scattering hypervolume is a three-body analog of the two-body $s$-wave or $p$-wave scattering length.

In this paper, we define the three-body scattering hypervolume $D_F$ of identical spin-polarized fermions in 2D, by studying the wave function of three such fermions colliding at zero energy and zero orbital angular momentum.
We find that in two dimensions $D_F$ has the dimension of length raised to the \emph{sixth} power, while in three dimensions its dimension is length raised to the \emph{eighth} power \cite{wang2021fermion3D}.

This article is organized as follows.
In Sec.~\ref{sec:asymp}, we first review the two-body wave functions for identical spin-polarized fermions in 2D, and then derive the asymptotic expansions of the wave function for three such fermions colliding at zero energy and zero orbital angular momentum; the parameter $D_F$ appears as a coefficient in these expansions.
In Sec.~\ref{sec:Born}, we derive an approximate formula for $D_F$ for weak interaction potentials using the Born expansion.
In Sec.~\ref{sec:energy}, we first consider three fermions in a large square with periodic boundary conditions
and derive the shifts of their energy eigenvalues due to a nonzero $D_F$, and then consider the dilute spin-polarized Fermi gas in 2D
and derive the shifts of its energy and pressure due to a nonzero $D_F$.
In Sec.~\ref{sec:recombination}, we study the dilute spin-polarized Fermi gas in 2D with interaction potentials that support
two-body bound states, for which we have three-body recombination processes and $D_F$ has nonzero imaginary part,
and derive formulas for the rates of these processes in terms of the imaginary part of $D_F$.
In Sec.~\ref{sec:summary} we summarize our results and discuss the generalization of the hypervolume to three-body collisions with
a higher orbital angular momentum.

\section{ASYMPTOTICS OF THE THREE-BODY WAVE FUNCTION\label{sec:asymp}}
We consider three identical spin-polarized fermions colliding with zero kinetic energy. Their wave function $\Psi$ satisfies the Schr\"{o}dinger equation
\begin{equation}\label{3body_equ}
	\sum_{i=1}^{3}\left[-\frac{\hbar^2}{2m}\nabla_{i}^2+ V(s_i)\right]\Psi+V_3(s_1,s_2,s_3)\Psi=0,
\end{equation}
where $m$ is the mass of each fermion, $\vect r_i$ is the position vector of the $i$th fermion, and $\vect s_i\equiv\vect r_j-\vect r_k$. The indices $(i,j,k)=(1,2,3)$, $(2,3,1)$, or $(3,1,2)$.
$V(s_i)$ is the two-body potential, and $V_{3}$ is the three-body potential.
We assume that the interactions among these fermions are finite-ranged, and depend only on the interparticle distances. 
We define the Jacobi coordinates for later use:
\begin{subequations}
	\begin{align}
		&\vect{R}_i \equiv\vect{r}_i-(\vect{r}_{j}+\vect{r}_{k})/2,\\
		&B\equiv\sqrt{\left(s_1^2+s_2^2+s_3^2\right)/2}=\sqrt{R_i^2+\frac34s_i^2},\\
		&\Theta_i \equiv\arctan\frac{2R_i}{\sqrt{3} s_i}.
	\end{align}
\end{subequations}
$B$ is called the hyperradius, and $\Theta_i$ is the hyperangle.

Equation~\eqref{3body_equ} and the translational invariance of $\Psi$ do not uniquely determine the wave function for the zero energy collision. We need to also specify the asymptotic behaviour of $\Psi$ when the three fermions are far apart. 
The leading-order term $\Psi_0$ in the wave function when $s_1,s_2,s_3$ go to infinity simultaneously should satisfy the Laplace equation $(\nabla_1^2+\nabla_2^2+\nabla_3^2)\Psi_0=0$ and scale as $B^p$ at large $B$.
The most important three-body wave function for zero-energy collisions, for purposes of understanding ultracold collisions, should be the one with the minimum value of $p$ \cite{wang2021fermion3D}. The reason is that the larger the value of $p$, the less likely for the three particles to come to the range of interaction within which they can interact. 
One can easily show that the minimum value of $p$ for three identical fermions in 2D is $p_{\text{min}}=2$, and
the corresponding $\Psi_0$ is
\begin{equation}
	\Psi_0=s_x R_y-s_y R_x,
\end{equation}
and it takes the same form as one of the $\Psi_0$'s in 3D;
here and in the following we define $\vect s\equiv\vect s_1$ and $\vect R\equiv\vect R_1$.
One can verify that $\Psi_0=s_{ix}R_{iy}-s_{iy}R_{ix}$ for any $i\in\{1,2,3\}$.
Unlike in 3D, however, we have only one linearly independent three-body wave function for the zero-energy collision with $p=2$ in 2D, and this wave function has \emph{zero} total orbital angular momentum and is rotationally invariant.

\subsection{Two-body special functions}
For two-body scattering in the center-of-mass frame with collision energy $E=\hbar^2 k^2/m$ and orbital angular momentum quantum number $l$, the wave function can be separated as $\psi(s,\theta)=u(s)e^{\pm i l \theta}/\sqrt{s}$, and the radial part $u(s)$ satisfies
\begin{equation}
	\frac{d^2 u}{ds^2}+\bigg[k^2-\frac{m V(s)}{\hbar^2}-\frac{l^2-1/4}{s^2}\bigg]u=0.
\end{equation}
We assume a finite range interaction such that $V(s)$ vanishes at $s>r_e$. The analytical formula for $u(s)$ at $s>r_e$ is
\begin{equation}
	u(s)=\alpha_l \sqrt{s}\big[J_l (ks)\cot \delta_l(k)-Y_l (ks)\big],
\end{equation}
where $J_l$ and $Y_l$ are the Bessel functions of the first and second kinds, respectively. $\alpha_l$ is an arbitrary coefficient which determines the overall amplitude of the two-body wave function. $\delta_l(k)$ is the $l$-wave scattering phase shift, and it satisfies the effective range expansion \cite{hammer2009causality,hammer2010causality}:
\begin{equation}
	k^{2l}\Big[\cot \delta_l(k)-\frac{2}{\pi}\ln (k\rho_l)\Big]=-\frac{1}{a_l}+\frac{1}{2}r_l k^2+O(k^4),
\end{equation}
where $a_l$ is the $l$-wave scattering ``length" ($l\geq1$) with dimension $[\mathrm{length}]^{2l}$.
$r_l$ is called the $l$-wave effective range for $l\neq1$ and it has dimension $[\mathrm{length}]^{2-2l}$.
$\rho_l$ is an arbitrary length scale. 

The wave function in the $l$-wave channel at $s>r_e$ is
\begin{equation}\label{two-body wave}
	\psi^{(l\pm)}(\vect{s})=-k^l a_l\big[J_l (ks)\cot\delta_l(k)-Y_l (ks)\big] e^{\pm i l \theta}.
\end{equation}
Here we have fixed the overall amplitude of $\psi^{(l\pm)}$ by specifying the coefficient $\alpha_l=-k^{l}a_l$.

At small collision energies, $k\ll 1/r_e$, the wave function can be expanded in powers of $k^2$ \cite{tan2008three,wang2021threebody,wang2021fermion3D}:
\begin{equation}\label{energyexpansion}
	\psi^{(l\pm)}(\vect s)=\phi^{(l\pm)}(\vect s)+k^2 f^{(l\pm)}(\vect s)+k^4 g^{(l\pm)}(\vect s)+\dots,
\end{equation}
where $\phi^{(l\pm)}(\vect s),f^{(l\pm)}(\vect s),g^{(l\pm)}(\vect s),\dots$ are called the two-body special functions, and they satisfy
\begin{subequations}
	\begin{align}
		&\widetilde{H}\phi^{(l\pm)}(\vect s)=0,\\
		&\widetilde{H}f^{(l\pm)}(\vect s)=\phi^{(l\pm)}(\vect s),\\
		&\widetilde{H} g^{(l\pm)}(\vect s)=f^{(l\pm)}(\vect s),\\
		&\dots,\nonumber
	\end{align}
\end{subequations}
where $\widetilde{H}$ is defined as 
\begin{equation}
	\widetilde{H} \equiv -\nabla_{\vect s}^2+ \frac{m}{\hbar^2} V(s).
\end{equation}
$\hbar^2\widetilde{H}/m$ is the two-body Hamiltonian for the collision of two fermions in the center-of-mass frame. The two-body special functions will appear in the expansion of the three-body wave function $\Psi$ when two fermions are held at a fixed distance
and the third fermion is far away from the two.

$\phi^{(l\pm)}(\vect s)$ is the wave function for the zero-energy collision of the two fermions in the $l$-wave channel.
From Eqs.~\eqref{two-body wave} and \eqref{energyexpansion} we get
\begin{equation}
	\phi^{(l\pm)}(\vect s)=\bigg[\frac{s^l}{(2l)!!}-\frac{2(2l-2)!!a_l}{\pi s^l}\bigg]e^{\pm i l\theta},~~s>r_e,\label{two-body-phi}
\end{equation}
for $l\geq1$. $l$ must be odd for identical spin-polarized fermions due to Fermi statistics. We use symbols $p,f,h,\cdots$ to represent $l=1,3,5,\cdots$, namely $a_1=a_p$, $r_1=r_p$, $a_3=a_f$, and so on.

For $f^{(l\pm)}(\vect{s})$ with $l=1$, we find
\begin{equation}
	f^{(1\pm)}(\vect s)=\Big(-\frac{1}{16}s^3+\frac{a_p s}{\pi} \ln \frac{s}{R_p}\Big)e^{\pm i \theta},~~s>r_e,
\end{equation}
where $R_p\equiv 2\rho_1 e^{\pi r_p/4+1/2-\gamma_{E}}$, and $\gamma_{E}=0.5772\cdots$ is Euler's constant. We call $R_p$ the $p$-wave effective range.

The explicit formulas for $f^{(l\pm)}$ with $l>1$ and $g^{(l\pm)}$ are not listed for brevity, as they are not used in this paper.

\subsection{The 111 expansion and the 21 expansion}
As what we did in previous works \cite{tan2008three,wang2021threebody,wang2021fermion3D}, we derive two asymptotic expansions for the three-body wave function $\Psi$.
When the three fermions are all far apart from each other, such that the pairwise distances $s_1$, $s_2$, $s_3$ go to infinity simultaneously for any fixed ratio $s_1:s_2:s_3$, we expand $\Psi$ in powers of $1/B$, and this expansion is called the 111 expansion.
When two fermions are held at a fixed distance $s$ but the third fermion is at large distance $R$ away from the center of mass of the two,
we expand $\Psi$ in powers of $1/R$, and this is called the 21 expansion. These expansions are
\begin{subequations}
	\begin{align}
		&\Psi=\sum_{p=-2}^{\infty} \mathcal{T}^{(-p)}(\vect{r}_1,\vect{r}_2,\vect{r}_3),\label{111-form}\\
		&\Psi=\sum_{q=-1}^{\infty}\mathcal{S}^{(-q)}(\vect{R},\vect{s}),\label{21-form}
	\end{align}
\end{subequations}
where $\vect R\equiv\vect R_1$, $\vect s\equiv\vect s_1$, $\mathcal{T}^{(-p)}$ scales as $B^{-p}$ and $\mathcal{S}^{(-q)}$ scales as $R^{-q}$. Here  $p$ starts from $p=-2$ because the leading order term $\Psi_0$ in the 111 expansion scales like $B^2$. Because $\Psi_0\propto R^1$ for any fixed $\vect s$, the leading order term in the 21 expansion should scale like $R^1$, so $q$ starts from $q=-1$.

$\mathcal{T}^{(-p)}$ satisfies the free Schr\"odinger equation outside of the interaction range:
\beq\label{free Schrodinger}
-\Big(\nabla_1^2+\nabla_2^2+\nabla_3^2\Big)\mathcal{T}^{(-p)}=0.
\eeq
If one fermion is far away from the other two, \Eq{3body_equ} becomes
\begin{equation}
	\Big(\widetilde{H}-\frac{3}{4}\nabla_{\vect{R}}^2\Big)\Psi =0.
\end{equation}
This leads to the following equations for $\mathcal{S}^{(-q)}$,
\begin{equation}
	\begin{split}
		&\widetilde{H} \mathcal{S}^{(1)}=0,~~\widetilde{H} \mathcal{S}^{(0)}=0,\\
		&\widetilde{H} \mathcal{S}^{(-q)}=\frac{3}{4}\nabla_{\vect{R}}^2 \mathcal{S} ^{(-q+2)}\quad (q\geq 1).
	\end{split}
\end{equation}

We can further expand $\mathcal{T}^{(-p)}$ as $\sum_{i} t^{(i,-p-i)}$ when $s\ll R$, where $t^{(i,j)}$ scales like $R^i s^j$.
We can also further expand $\mathcal{S}^{(-q)}$ as $\sum_{j} t^{(-q,j)}$ when $s\gg r_e$. 
Because the three-body wave function $\Psi$ may be expanded as $\sum_p \mathcal{T}^{(-p)}$ at $B\to\infty$,
and may also be expanded as $\sum_q \mathcal{S}^{(-q)}$ at $R\to\infty$, the $t^{(i,j)}$ in the above two expansions
should be the same. Actually the wave function has a double expansion $\Psi=\sum_{i,j}t^{(i,j)}$ in the region $r_e\ll s\ll R$.

\begin{figure}[htb]
	\includegraphics[width=0.5\textwidth,height=0.5\textwidth]{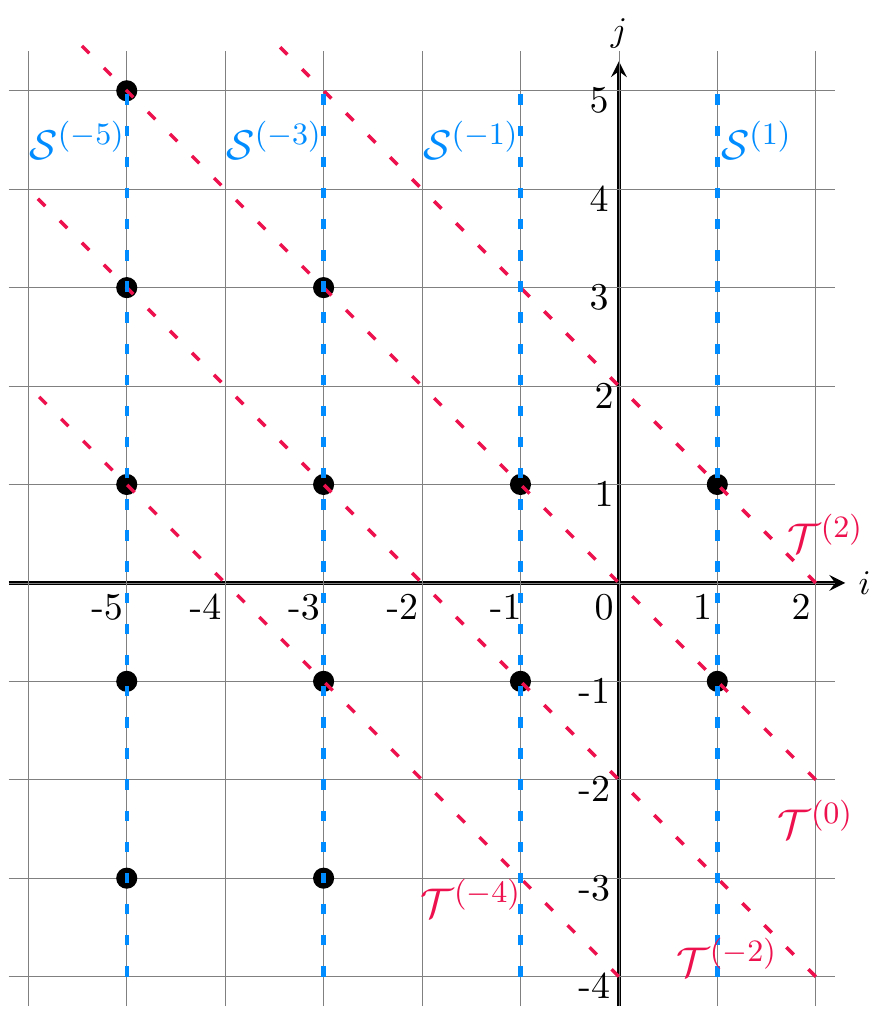}
	\caption{\label{fig:expansion} Diagram of the points representing $t^{(i,j)}$ on the $(i,j)$ plane. Each point with coordinates $(i,j)$ represents $t^{(i,j)}$ which scales like $R^i s^j$. Thick dots represent those points at which $t^{(i,j)}\ne0$.
		The term $\mathcal{T}^{(-p)}$ in the 111 expansion is represented by a red dashed line satisfying the equation $i+j=-p$. The term $\mathcal{S}^{(-q)}$ in the 21 expansion is represented by a blue dashed line satisfying the equation $i=-q$.}
\end{figure}

We show the points at which $t^{(i,j)}\ne0$ on the $(i,j)$ plane in Fig.~\ref{fig:expansion}. $\mathcal{T}^{(-p)}$ corresponds to 
a straight line with slope equal to $-1$ and intercept equal to $-p$. $\mathcal{S}^{(-q)}$ corresponds to a vertical line $i=-q$ in the diagram. 

To derive the two expansions, we start from the leading-order term (which fixes the overall amplitude of $\Psi$) in the 111 expansion:
\begin{equation}
	\mathcal{T}^{(2)}=\Psi_0=(s_x R_y-s_y R_x).
\end{equation}
We then first derive $\mathcal{S}^{(1)}$, and then derive $\mathcal{T}^{(1)}$, and then derive $\mathcal{S}^{(0)}$, and then derive $\mathcal{T}^{(0)}$, and so on, all the way until $\mathcal{S}^{(-5)}$. At every step, we require the 111 expansion and the 21 expansion to be consistent in the region $r_e\ll s\ll R$. 
See appendix \ref{appendix:derivation} for the details of derivation.

The resultant 111 expansion is
\begin{widetext}
	\begin{equation}\label{111}
		\begin{split}
			\Psi=(s_x R_y-s_y R_x) &\bigg\lbrace 
			1-\frac{3D_F}{2\pi^2 B^6}-\frac{4 a_p}{\pi}\sum_{i=1}^{3} \frac{1}{s_i^2}
			+\frac{32a_p^2}{\pi^2 B^2}\sum_{i=1}^{3}\frac{1}{s_i^2}
			-\frac{160a_p^3}{\pi^3 B^6}\sum_{i=1}^3\bigg(\frac{R_i^2}{s_i^2}+3\ln\frac{B^2}{\sqrt{|a_p|}s_i}\bigg) +O(B^{-7})\bigg\rbrace,
		\end{split}
	\end{equation}
	where $D_F$ is the three-body scattering hypervolume of identical spin-polarized fermions in 2D, and it appears at the order of $B^{-4}$ in the expansion of $\Psi$. The dimension of $D_F$ is length raised to the sixth power.
	
	The resultant 21 expansion  is
	\begin{equation}\label{21}
		\begin{split}
			\Psi=&i \left( R-\frac{8a_p}{\pi R}+\frac{40a_p^2}{\pi^2 R^3} -\frac{\xi}{R^5}-\frac{1920a_p^3}{\pi^3 R^5}\ln \frac{R}{\widetilde{R}}\right) \Big[y^{(1-)}(\widehat{\vect{R}})\phi^{(1+)}(\vect{s})-y^{(1+)}(\widehat{\vect{R}})\phi^{(1-)}(\vect{s})\Big] \\
			&+i \left(-\frac{48a_p}{\pi R^3}+\frac{384a_p^2}{\pi^2 R^5}\right)\Big[y^{(3-)}(\widehat{\vect{R}})\phi^{(3+)}(\vect{s})-y^{(3+)}(\widehat{\vect{R}})\phi^{(3-)}(\vect{s})\Big]\\
			&-\frac{960i a_p}{\pi R^5} \Big[y^{(5-)}(\widehat{\vect{R}})\phi^{(5+)}(\vect{s})-y^{(5+)}(\widehat{\vect{R}})\phi^{(5-)}(\vect{s})\Big]
			+\frac{240i a_p^2}{\pi^2 R^5}\Big[y^{(1-)}(\widehat{\vect{R}})f^{(1+)}(\vect{s})-y^{(1+)}(\widehat{\vect{R}})f^{(1-)}(\vect{s})\Big]+O(R^{-6}),
		\end{split}
	\end{equation}
\end{widetext}
where
\begin{align}
	&y^{(l\pm)}(\widehat{\vect{R}})\equiv (R_x\pm i R_y)^l/R^l,\\
	&\widetilde{R} = |a_p|^{3/8}R_p^{1/4},\\
	&\xi=\frac{3D_F}{2\pi^2}-\frac{280 a_p^3}{\pi^3}.
\end{align}

For any finite-range potentials $V(s)$ and $V_3(s_1,s_2,s_3)$, we may solve \Eq{3body_equ} numerically and match the solution to the asymptotic expansions in Eqs.~\eqref{111} and \eqref{21} at large inter-fermionic distances to determine $D_F$ numerically.
But if the potentials are sufficiently weak, we may use the Born expansion to calculate $D_F$, as we do in Sec.~\ref{sec:Born}.

For attractive potentials whose strengths are fine-tuned such that there is a three-body $S$-wave bound state (ie, with total orbital angular momentum quantum number $L=0$) with energy close to zero,
we anticipate that $D_F$ becomes large;
when we tune the strength of attraction such that this shallow three-body bound state barely forms (so that there is a metastable
three-body state whose energy is positive and has some width due to the finite lifetime), $D_F$ is large and negative;
when we increase the strength of attraction such that this shallow three-body bound state has a small negative energy,
$D_F$ is large and positive; $D_F$ has a pole as a function of the strength of attraction, and the pole is located
at the critical strength at which the energy of this three-body bound state is zero.

\section{Approximate formula of $D_F$ for weak interaction potentials}\label{sec:Born}
For weak interactions, we analytically derive an approximate formula of $D_F$ by using the Born series:
\begin{equation}
	\Psi=\Psi_0+\widehat{G}\mathcal{V}\Psi_0+(\widehat{G}\mathcal{V})^2\Psi_0+\cdots
\end{equation}
where 
$\mathcal{V}$ is the total interaction potential and $\widehat{G}$ is the Green's operator ($\widehat{G}=-\widehat{H}_0^{-1}$, $\widehat{H}_0$ is the three-body kinetic energy operator). Writing the interaction potential as $\mathcal{V}=V_3(s_1,s_2,s_3)+\sum_{i=1}^3 V(s_i)$, we find
\begin{subequations}
	\begin{align}
		&\widehat{G}\mathcal{V}\Psi_0=-(s_x R_y-s_y R_x)\left(\frac{6\Lambda }{\pi B^6}+\sum_{i=1}^3 \frac{\alpha_2}{2 s_i^2}+\dots\right),\\
		&(\widehat{G}\mathcal{V})^2\Psi_0=(s_x R_y-s_y R_x)\nonumber\\
		&\quad\quad\quad\times\bigg\{\Big[\sum_{i=1}^3 \left( \frac{\beta_2}{2 s_i^2}+\frac{\alpha_2^2}{2B^2 s_i^2}\right)\Big] -\frac{9\alpha_2\alpha_4}{16 B^6}+\dots\bigg\}\nonumber\\
		&\quad\quad\quad+O(VV_3)+O(V_3^2),
	\end{align}
\end{subequations}
where
\begin{subequations}
	\begin{align}
		&\alpha_n\equiv \frac{m}{\hbar^2}\int_{0}^{\infty}\!\!\! ds~ s^{n+1}V(s),\\
		&\beta_2\equiv \frac{m^2}{\hbar^4}\int_{0}^{\infty}\!\!\! ds\int_{0}^{s}\!\!\! ds' ~s s'^3 V(s) V(s'),\\
		&\Lambda\equiv \frac{m}{\hbar^2}\int ds_1 ds_2 ds_3~s_1 s_2 s_3 V_3(s_1, s_2, s_3) S_{\Delta}(s_1,s_2,s_3).\label{Lambda}
	\end{align}
\end{subequations}
$S_{\Delta}=\sqrt{p(p-s_1)(p-s_2)(p-s_3)}$ is the area of the triangle with sides $s_1,s_2,s_3$, where $p=(s_1+s_2+s_3)/2$.
The integral on the right hand side of \Eq{Lambda} is over all values of $s_1,s_2,s_3$ satisfying $s_1>0$, $s_2>0$, $s_3>0$, $s_1+s_2>s_3$, $s_2+s_3>s_1$, and $s_3+s_1>s_2$ simultaneously.
The details of the derivation are shown in appendix \ref{appendix:Born}.

By comparing these results with the 111 expansion in \Eq{111}, we find the expansions of $a_p$ and $D_F$ in powers of $V(s)$ and $V_3(s_1,s_2,s_3)$:
\begin{align}
	&a_p=\frac{\pi}{8}(\alpha_2-\beta_2)+O(V^3),\label{ap-weak}\\
	&D_F=4\pi \Lambda+\frac{3\pi^2}{8}\alpha_2\alpha_4+O(VV_3)+O(V_3^2)+O(V^3)\label{DF-weak}.
\end{align}
For any particular two-body potential $V(s)$, e.g. the square-well potential, one can calculate $a_p$ analytically and verify that the result is consistent with \Eq{ap-weak} if $V$ is weak. 
Equation \eqref{DF-weak} shows that $D_F$ is quadratically dependent on the two-body potential $V$ if $V$ is weak
and the three-body potential $V_3$ is absent.
On the other hand, $D_F$ is linearly dependent on the three-body potential $V_3$ if $V_3$ is weak and the two-body potential is absent.

If the interactions are not weak, one can solve the three-body Schr\"{o}dinger equation numerically at zero energy and zero orbital angular momentum and match the resultant wave function with the asymptotic expansion in \Eq{111} or \Eq{21} to numerically extract the value of $D_F$.

\section{Shifts of the energy and the pressure of identical spin-polarized fermions}\label{sec:energy}

In this section, we first study the energy shift of three spin-polarized fermions caused by the scattering hypervolume $D_F$. 
We then derive the shifts of the thermodynamic properties, including the energy and the pressure, of the spin-polarized Fermi gas due to a nonzero $D_F$.

\subsection{Three fermions in a large square}
For the sake of simplicity, in this section we assume that the fermions have no two-body interaction or have a fine-tuned two-body interaction such that the two-body $p$-wave scattering ``length" $a_p=0$ but the three-body scattering hypervolume $D_F\ne0$, and the $111$ expansion for the zero-energy three-body wave function in \Eq{111} is simplified as
\beq
\Psi \simeq
(s_x R_y-s_y R_x)
\left(1-\frac{3D_F}{2\pi^2 B^6}\right).
\eeq
For the purpose of calculating the energy shifts due to a nonzero $D_F$, the true interaction potential $V(s_1)+V(s_2)+V(s_3)+V_{3}(s_1,s_2,s_3)$, which in general has a complicated dependence on the interparticle distances, is replaced by a three-body pseudopotential $V_{\text{ps}}$.
We use the following pseudopotential,
\begin{equation}\label{Vps}
	V_{\text{ps}}=\frac{\hbar^2D_F}{6m}\Big\{\Big[\nabla_{\vect{s}}^2\nabla_{\vect{R}}^2-(\nabla_{\vect{s}}\cdot\nabla_{\vect{R}})^2\Big]\delta(\vect{s})\delta(\vect{R})\Big\}\Lambda,
\end{equation}
where $\Lambda$ is a projection operator which, when acting on the $O(B^{-4})$ term in the three-body wave function, yields zero.
The operator $\Lambda$ is an analog of the operator $\frac{\partial}{\partial r}r$ in the two-body pseudopotential for $s$-wave  collisions in Refs.~\cite{huang1957quantum,lee1957eigenvalues}.
One can check the pseudopotential in \Eq{Vps} is symmetric under the interchange of the three fermions. The coefficient on the right hand side of \Eq{Vps} has been chosen such that
\begin{equation}
	-\frac{\hbar^2}{2m}(\nabla_1^2+\nabla_2^2+\nabla_3^2)\Psi+V_{\textrm{ps}}\Psi=0.
\end{equation}

We now consider three fermions in a large square with area $A$, and impose the periodic boundary conditions on the wave function. Consider an energy eigenstate in which the momenta of the fermions are $\hbar\vect{k}_1$, $\hbar\vect{k}_2$ and $\hbar\vect{k}_3$ in the absence of interactions.
When we introduce interactions that give rise to a nonzero $D_F$, the energy eigenvalue of the three-body state is shifted
by the following amount at first order in the perturbation:
\begin{equation}
	\mathcal{E}_{\vect{k}_1\vect{k}_2\vect{k}_3}=\int d^2 \vect{r}_1 d^2\vect{r}_2 d^2\vect{r}_3\, |\Psi_{\vect k_1\vect k_2\vect k_3}|^2 V_{\text{ps}},
\end{equation}
where $\Psi_{\vect k_1\vect k_2\vect k_3}$ is the normalized unperturbated wave function and it can be written in terms of a Slater determinant:
\begin{equation}
	\Psi_{\vect k_1\vect k_2\vect k_3}=\frac{1}{\sqrt{6}A^{3/2}}
	\left| \begin{matrix}
		e^{i \vect{k}_1\cdot\vect{r}_1}& e^{i \vect{k}_1\cdot\vect{r}_2} & e^{i \vect{k}_1\cdot\vect{r}_3}\\
		e^{i \vect{k}_2\cdot\vect{r}_1}& e^{i \vect{k}_2\cdot\vect{r}_2} & e^{i \vect{k}_2\cdot\vect{r}_3}  \\ 
		e^{i \vect{k}_3\cdot\vect{r}_1}& e^{i \vect{k}_3\cdot\vect{r}_2} & e^{i \vect{k}_3\cdot\vect{r}_3}
	\end{matrix}\right|  .
\end{equation}
We get
\begin{equation}
	\mathcal{E}_{\vect{k}_1\vect{k}_2\vect{k}_3}=\frac{\hbar^2 D_F}{3m A^2}\left(\vect{k}_1\times\vect{k}_2+\vect{k}_2\times\vect{k}_3+\vect{k}_3\times\vect{k}_1\right)^2.\label{E of 3 fermions}
\end{equation}
This energy shift is proportional to the square of the area of the $\vect{k}$-space triangle whose vertices are $\vect k_1$, $\vect k_2$, and $\vect k_3$.

\subsection{Energy shift of many fermions and thermodynamic consequences}

We generalize the energy shift to $N$ fermions in the periodic area $A$. The number density of the fermions is $n=N/A$.
We define the Fermi wave number $k_F=(4\pi n)^{1/2}$, the Fermi energy $\epsilon_F=\hbar^2 k_F^2/2m$, and the Fermi temperature $T_F=\epsilon_F/k_B$, where $k_B$ is the Boltzmann constant. 

\subsubsection{Adiabatic shifts of energy and pressure in the thermodynamic limit}
Starting from a many-body state at a finite temperature $T$, if we introduce a nonzero $D_F$ \emph{adiabatically}, the energy shift at first order in $D_F$ is equal to the sum of the contributions from all the triplets of fermions, namely
\begin{equation}
	\Delta E=\frac{1}{6}\sum_{\vect{k}_1\vect{k}_2\vect{k}_3}\mathcal{E}_{\vect{k}_1\vect{k}_2\vect{k}_3}\, n_{\vect{k}_1}n_{\vect{k}_2}n_{\vect{k}_3},
\end{equation}
where $n_{\vect{k}}=(e^{\beta(\epsilon_{\vect{k}}-\mu)}+1)^{-1}$ is the Fermi-Dirac distribution function, $\beta=1/k_B T$, $\epsilon_{\vect{k}}=\hbar^2 k^2/2m$ is the kinetic energy of a fermion with linear momentum $\hbar\vect k$, and $\mu$ is the chemical potential. The summation over $\vect{k}$ can be replaced by a continuous integral $\sum_{\vect{k}}=A \int d^2k/(2\pi)^2$ in the thermodynamic limit. Carrying out the integral, we get
\begin{equation}
	\Delta E(T)=\frac{N\hbar^2 D_F}{192\pi^2 m}k_F^{8}\widetilde{T}^4  [f_{2}(e^{\beta\mu})]^2,
\end{equation}
where $\widetilde{T}=T/T_F$, and the function $f_{\nu}(z)$ is defined as
\begin{equation}
	f_{\nu}(z)\equiv -\mathrm{Li}_{\nu}(-z)=\frac{2}{\Gamma(\nu)}\int_0^{\infty}\!\! dx~ \frac{x^{2\nu-1}}{1+e^{x^2}/z}.
\end{equation}
The number of fermions satisfies
$
N=\sum_{\vect{k}} \frac{1}{e^{\beta (\epsilon_{\vect{k}}-\mu)}+1}
$,
and this leads to
\begin{equation}
	\widetilde{\mu}=\widetilde{T}\ln \left(e^{1/\widetilde{T}}-1\right),
\end{equation}
where $\widetilde{\mu}=\mu/\epsilon_F$.

\begin{figure*}[htb]
	\centering   	
	\subfloat[Energy]
	{
		\begin{minipage}[t]{0.5\textwidth}
			\includegraphics[width=1\textwidth]{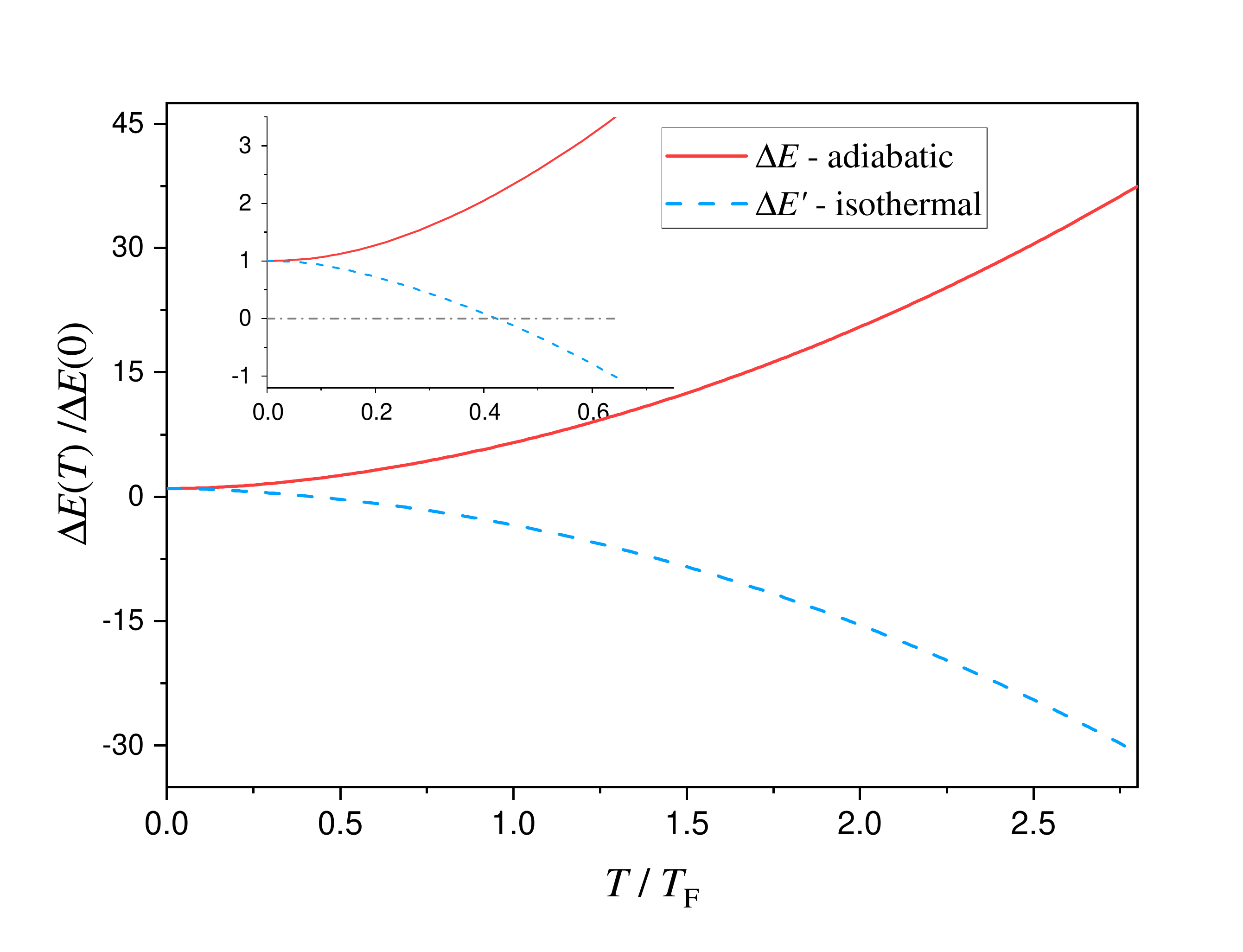}   
		\end{minipage}%
	}
	\subfloat[Pressure]
	{
		\begin{minipage}[t]{0.5\textwidth}
			\includegraphics[width=1\textwidth]{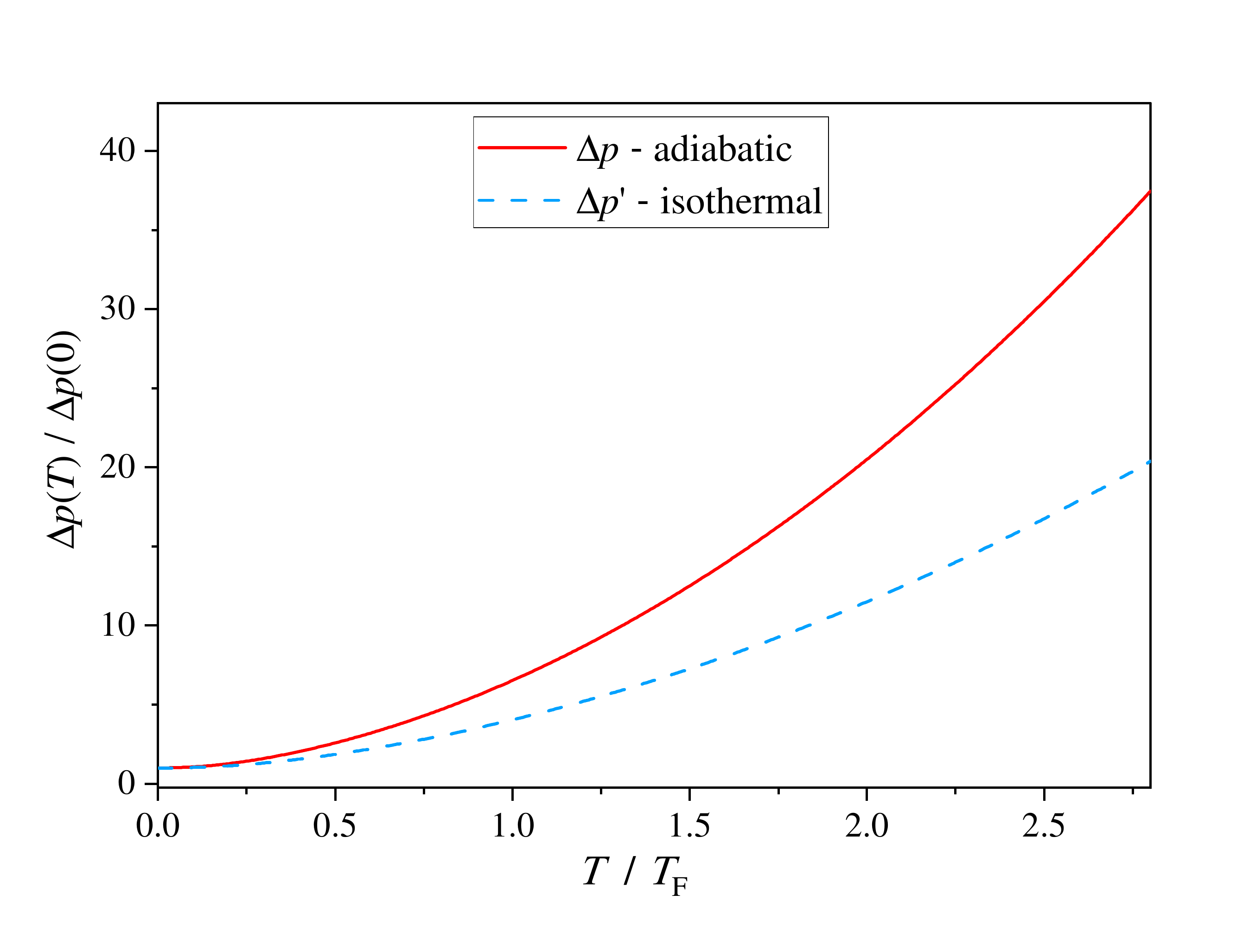} 
		\end{minipage}
	}	
	\caption{The shifts of energy (a) and pressure (b) caused by the adiabatic (red line) or isothermal (blue dashed line) introduction of $D_F$ vs the  temperature $T$. At $T\simeq0.424T_F$, the isothermal energy shift $\Delta E$ changes sign.}
	\label{fig:energy and pressure}
\end{figure*}

In the low temperature limit, $T\ll T_F$, 
\begin{align}
	\Delta E(T)=&\frac{N\hbar^2 D_F}{192\pi^2 m}k_F^{8}\nonumber\\
	\times&\Big[\frac{1}{4}+\frac{\pi^2}{6}\widetilde{T}^2+\frac{\pi^4}{36}\widetilde{T}^4+O(\widetilde{T}e^{-1/\widetilde{T}})\Big].
\end{align}
In an intermediate temperature regime, $T_F\ll T\ll T_e$, 
\begin{equation}
	\Delta E(T)=\frac{N\hbar^2 D_F}{192\pi^2 m}k_F^{8}
	\left[\widetilde{T}^2+\frac{\widetilde{T}}{2}+\frac{17}{144}+O(\widetilde{T}^{-1})\right],
\end{equation}
where $T_e=\frac{\hbar^2}{2mr_e^2k_B}$.
If $T$ is comparable to or higher than $T_e$, the de Broglie wave lengths of the fermions will be comparable to or shorter than the range $r_e$ of interparticle interaction potentials, and we can no longer use the effective parameter $D_F$ to describe the system.
See Fig. \ref{fig:energy and pressure}~(a) for $\Delta E$ as a function of the initial temperature.

The pressure of the spin-polarized Fermi gas changes by the following amount due to the adiabatic introduction of $D_F$:
\begin{equation}
	\begin{split}
		\Delta p&=-\left(\frac{\partial \Delta E}{\partial A}\right)_{S,N}
		=\frac{4\Delta E}{A}\\
		&=\frac{n\hbar^2 D_F}{48\pi^2 m}k_F^{8}\widetilde{T}^4  [f_{2}(e^{\beta\mu})]^2.\label{pressure adia}
	\end{split}
\end{equation}
The subscripts $S,N$ in \Eq{pressure adia} mean that we keep the entropy $S$ and particle number $N$ fixed when taking the partial derivative. See Fig. \ref{fig:energy and pressure}~(b) for $\Delta p$ as a function of the initial temperature.

\subsubsection{Isothermal shifts of energy and pressure in the thermodynamic limit}
If the interaction is introduced adiabaticly, the temperature will increase (if $D_F>0$) or decrease (if $D_F<0$). The change of temperature is
\begin{equation}
	\Delta T=\left(\frac{\partial \Delta E}{\partial S}\right)_{N,A}.
\end{equation}
Therefore if we introduce $D_F$ isothermally, the energy shift $\Delta E'$ should be
\begin{equation}
	\Delta E'=\Delta E-C \Delta T=\left(1-T \frac{\partial}{\partial T}\right)\Delta E,
\end{equation}
where $C$ is the heat capacity of the noninteracting Fermi gas at constant volume.
In the low temperature limit, $T\ll T_F$,
\begin{equation}\label{DeltaE'lowT}
	\Delta E'(T)=\frac{N\hbar^2 D_F}{192\pi^2 m}k_F^{8}
	\Big[\frac{1}{4}-\frac{\pi^2}{6}\widetilde{T}^2-\frac{\pi^4}{12}\widetilde{T}^4
	+O(e^{-1/\widetilde{T}})\Big].
\end{equation}
In an intermediate temperature regime, $T_F\ll T\ll T_e$, 
\begin{equation}\label{DeltaE'highT}
	\Delta E'(T)=\frac{N\hbar^2 D_F}{192\pi^2 m}k_F^{8}
	\Big[-\widetilde{T}^2+\frac{17}{144}+O(\widetilde{T}^{-1})\Big].
\end{equation}
According to Eqs.~\eqref{DeltaE'lowT} and \eqref{DeltaE'highT}, $\Delta E'$ changes sign as we increase the temperature from $T\ll T_F$ to $T\gg T_F$.
Therefore, there is a critical temperature $T_c$ at which $\Delta E'=0$. We find 
\begin{equation}
	T_c \simeq 0.424 T_F.
\end{equation}

The pressure of the spin-polarized Fermi gas changes by the following amount due to the isothermal introduction of $D_F$:
\begin{equation}
	\Delta p'=\Delta p-\frac{ C\Delta T}{A}=
	\left(1-\frac{1}{4}T \frac{\partial}{\partial T}\right)\Delta p.
\end{equation}
In the low temperature limit, $T\ll T_F$,
\begin{equation}
	\Delta p'=\frac{n\hbar^2 D_F}{96\pi^2 m}k_F^{8}
	\left[\frac{1}{2}+\frac{\pi^2}{6}\widetilde{T}^2+O(e^{-1/\widetilde{T}})\right].
\end{equation}
In an intermediate temperature regime, $T_F\ll T\ll T_e$,
\begin{equation}
	\Delta p'=\frac{n\hbar^2 D_F}{96\pi^2 m}k_F^{8}
	\left[\widetilde{T}^2+\frac{3}{4}\widetilde{T}+\frac{17}{72}+O(\widetilde{T}^{-1})\right],
\end{equation}

The shifts of energy and pressure are plotted as functions of temperature in Fig.~\ref{fig:energy and pressure}~(a) and Fig.~\ref{fig:energy and pressure}~(b) respectively.


\section{The Three-body recombination rate\label{sec:recombination}}
If the collision of the three particles is purely elastic, $D_F$ is a real number.
But if the two-body interactions support bound states, then the three-body collisions are usually not purely elastic, and the three-body recombination may happen, and this is the case for most ultracold atomic gases because most ultracold atoms have two-body bound states. In this case $D_F$ becomes complex and acquires a \emph{negative} imaginary part, and the three-body recombination rate constant is proportional to the imaginary part of $D_F$ \cite{zhu2017threebody,braaten2006universality}. When the imaginary part of $D_F$ is nonzero,
the 111 expansion in \Eq{111} remains valid, but the 21 expansion \Eq{21} should be modified by
including terms describing bound pairs flying apart from the third particle.

Within a short time $\Delta t$, the probability that no recombination occurs is $\mathrm{exp}(-2|\mathrm{Im} {E}|\Delta t/\hbar)\simeq 1-2|\mathrm{Im} {E}|\Delta t/\hbar$. Then the probability of one recombination event is $2|\mathrm{Im} {E}|\Delta t/\hbar$.
Since each recombination event causes the loss of three low-energy fermions, the change of the number of remaining low-energy fermions in the short time $dt$ is
\begin{equation}
	d N=-\frac{1}{6}\sum_{\vect{k}_1\vect{k}_2\vect{k}_3}3\cdot\frac{2\D t}{\hbar}|\mathrm{Im}\mathcal{E}_{\vect{k}_1\vect{k}_2\vect{k}_3}| n_{\vect{k}_1}n_{\vect{k}_2}n_{\vect{k}_3}.
\end{equation}
This leads to
\begin{equation}
	\frac{d n}{d t}=-L_3 n^{3},
\end{equation}
and the coefficient $L_3$ is
\begin{equation}
	L_3=\frac{1}{2}\widetilde{T}^4 [f_{2}(e^{\beta\mu})]^2 \frac{\hbar|\mathrm{Im} D_F|}{m}k_F^4.
\end{equation}
$L_3$ depends on the density $n$ and the temperature $T$.

In the low temperature limit, $T\ll T_F$,
\begin{equation}
	L_3\simeq \frac{1}{8}\left(1+\frac{2\pi^2}{3}\widetilde{T}^2\right)\frac{\hbar|\mathrm{Im}D_F|}{m}k_F^4.
\end{equation}
In particular, at $T=0$,
\begin{equation}
	L_3=\frac{\hbar|\mathrm{Im} D_F|}{8m} k_F^4,
\end{equation}
and $L_3$ is proportional to $n^2$, so $d n/d t$ is proportional to $n^5$.

In an intermediate temperature regime, $T_F\ll T\ll T_e$, we find that
\begin{equation}\label{L3highT}
	L_3\simeq \frac{2m}{\hbar^3}|\mathrm{Im}D_F|(k_B T)^2.
\end{equation}
In this case $L_3$ is proportional to $T^2$, and is approximately independent of $n$, so $d n/d t$ is proportional to $n^3$. 
In Refs.~\cite{PhysRevA.65.010705,PhysRevA.90.042707} it was predicted that $L_3\propto T^2$ for the spin-polarized Fermi gas in three or two  dimensions according to the threshold law, and our \Eq{L3highT} is consistent with this prediction.
In Ref.~\cite{PhysRevA.99.052704} the $T^2$ dependence of $L_3$ was experimentally confirmed for a two-dimensional spin-polarized Fermi gas. One can extract the imaginary part of $D_F$ from the experimental value of $L_3$, by using the formulas we have presented here.

\section{Summary and Discussion\label{sec:summary}}
We have derived the asymptotic expansions of the three-body wave function $\Psi$ for identical spin-polarized fermions colliding at zero energy and zero orbital angular momentum in two dimensions. 
The scattering hypervolume $D_F$ appears at the order of $B^{-4}$ in the 111 expansion of $\Psi$.
We find that in two dimensions $D_F$ has the dimension of length raised to the \emph{sixth} power. In contrast, the dimension of $D_F$ for identical spin-polarized fermions in three dimensions is length raised to the \emph{eighth} power \cite{wang2021fermion3D}.

For weak interaction potentials, we have derived an approximate formula for $D_F$ by using the Born expansion.
For stronger interactions, one can solve the three-body Schr\"{o}dinger equation numerically at zero energy and zero orbital angular momentum and match the resultant wave function with the asymptotic expansion formulas we have derived in this paper to numerically extract the values of $D_F$.

We considered three fermions in a large square with periodic boundary conditions
and derived the shifts of their energy eigenvalues due to a nonzero $D_F$, and then considered the dilute spin-polarized Fermi gas in 2D and derived the shifts of its energy and pressure due to a nonzero $D_F$.

Finally, we studied the dilute spin-polarized Fermi gas in 2D with interaction potentials that support two-body bound states, for which we have three-body recombination processes and $D_F$ has nonzero imaginary part,
and we derived formulas for the three-body recombination rate constant $L_3$ in terms of the imaginary part of $D_F$ and the temperature and density of the Fermi gas.

We emphasize that the scattering hypervolume $D_F$ we have studied in this work refers to the $S$-wave collisions of three identical spin-polarized fermions in 2D, ie. the total orbital angular momentum quantum number $L$ is zero for such collisions.
One can also study the $P$-wave collisions of these three fermions in 2D, with $L=1$, and define a new three-body scattering hypervolume $D_F^{(P)}$, by studying the asymptotic expansions of the $P$-wave three-body wave function at zero collision energy.
The $P$-wave three-body collisions are interesting because there is super Efimov effect for $L=1$ near two-body $p$-wave resonances \cite{superefimov,Volosniev_2014,Gridnev2014,gaochao2015}.
The behavior of $D_F^{(P)}$ near such a resonance is expected to be related to the super Efimov effect and deserves future investigation.

\begin{acknowledgments}
	This work was supported by the National Key R\&D Program of China (Grants No.~2019YFA0308403) and National Key R\&D Program of China (Grant Number 2021YFA1400902).
\end{acknowledgments}
	
\appendix
\section{Derivation of the 111 expansion and the 21 expansion}\label{appendix:derivation}
We expand the three-body wave function in two forms:
\begin{subequations}
	\begin{align}
		&\Psi=\sum_{p=-2}^{\infty} \mathcal{T}^{(-p)}(\vect{r}_1,\vect{r}_2,\vect{r}_3),\label{a111-form}\\
		&\Psi=\sum_{q=-1}^{\infty}\mathcal{S}^{(-q)}(\vect{R},\vect{s}),\label{a21-form}
	\end{align}
\end{subequations}
where $\mathcal{T}^{(-p)}$ scales as $B^{-p}$, $\mathcal{S}^{(-q)}$ scales as $R^{-q}$. The hyperradius $B$ and the vectors $\vect{R}$ and $\vect{s}$ are already defined in the main text. 

If $s\ll R$, we can further expand $\mathcal{T}^{(-p)}$ as
\beq
\mathcal{T}^{(-p)}=\sum_{i} t^{(i,-p-i)},\label{T-p}
\eeq
where $t^{(i,j)}$ scales as $R^i s^j$.
If $s\gg r_e$, we can expand $\mathcal{S}^{(-q)}$ as
\beq
S^{(-q)}=\sum_{j} t^{(-q,j)}.
\eeq
Because the three-body wave function $\Psi$ may be expanded as $\sum_p \mathcal{T}^{(-p)}$ at $B\to\infty$,
and may also be expanded as $\sum_q \mathcal{S}^{(-q)}$ at $R\to\infty$, the $t^{(i,j)}$ in the above two expansions should be the same. In fact the wave function has a double expansion $\Psi=\sum_{i,j}t^{(i,j)}$ in the region $r_e\ll s\ll R$.

\paragraph*{\textbf{Step 1.}}
We start from the leading-order term in the 111 expansion:
\begin{equation}
	\mathcal{T}^{(2)}=s_x R_y-s_y R_x=t^{(1,1)},\label{t11}
\end{equation}
and this indicates that $\mathcal{S}^{(1)}$ is nonzero, but $\mathcal{S}^{(2)}$, $\mathcal{S}^{(3)}$, $\mathcal{S}^{(4)}$, \dots are zero.
Consequently
\beq
t^{(i,j)}=0,~~\text{if}~i\ge2.
\eeq
Expanding $\mathcal{T}^{(2)}$ at $s\ll R$, we find that
\beq
t^{(0,2)}=t^{(-1,3)}=t^{(-2,4)}=t^{(-3,5)}=\cdots=0.
\eeq
Since $\mathcal{T}^{(3)}$, $\mathcal{T}^{(4)}$, $\mathcal{T}^{(5)}$, \dots are zero, we have
\beq
t^{(i,j)}=0,~~\text{if}~i+j\ge3.
\eeq

\paragraph*{\textbf{Step 2.}}
At $s\gg r_e$ we expand $\mathcal{S}^{(1)}$ as
\begin{equation}\label{S1expand}
	\mathcal{S}^{(1)}=t^{(1,1)}+\sum_{j\le0}t^{(1,j)}.
\end{equation}
$\mathcal{S}^{(1)}$ also satisfies
\begin{equation}
	\widetilde{H} \mathcal{S}^{(1)}=0,
\end{equation}
where $\widetilde{H}$ is proportional to the two-body Hamiltonian, and has been defined in the main text. Therefore, $\mathcal{S}^{(1)}$ takes the following form
\begin{equation}
	\mathcal{S}^{(1)}=R\sum_l\Big[c_{l+} \phi^{(l+)}(\vect{s})+c_{l-}\phi^{(l-)}(\vect{s})\Big]. \label{S1}
\end{equation}
Because $\phi ^{(l\pm)}$ contributes a term proportional to $s^l$, $S^{(1)}$ contains a term scaling as $R^1 s^l$. On the other hand, the leading order term on the right hand side of \Eq{S1expand} is $t^{(1,1)}$ which scales as $R^1s^1$. So we must have $c_{l\pm}=0$ for $l>1$.

Expanding \Eq{S1} at $s\gg r_e$ to the order $s^1$, and using \Eq{two-body-phi}, we get
\begin{equation}
	t^{(1,1)}=\frac{1}{2}R[c_{1+}(s_x+\I s_y)+c_{1-}(s_x-\I s_y)].
\end{equation}
Comparing this result with \Eq{t11}, we find the coefficients $c_{1\pm}$:
\begin{subequations}
	\begin{align}
		&c_{1+}=\frac{\I}{R} (R_x-\I R_y),\\
		&c_{1-}=\frac{-\I}{R} (R_x+\I R_y).
	\end{align}
\end{subequations}
Therefore,
\begin{equation}
	\mathcal{S}^{(1)}=\I R\Big[y^{(1-)}(\hat{\vect{R}})\phi^{(1+)}(\vect{s})-y^{(1+)}(\hat{\vect{R}})\phi^{(1-)}(\vect{s})\Big] .
\end{equation}
Expanding $\mathcal{S}^{(1)}$ at $s\gg r_e$, we get
\begin{subequations}
	\begin{align}
		&t^{(1,0)}=0,\\
		&t^{(1,-1)}=-\frac{4a_p}{\pi s^2}(s_xR_y-s_y R_x),\label{t1,-1}\\
		&t^{(1,j)}=0,~~j\le-2.
	\end{align}
\end{subequations}

\paragraph*{\textbf{Step 3.}}
At $s\ll R$ we expand $\mathcal{T}^{(1)}$ as
\beq
\mathcal{T}^{(1)}=t^{(1,0)}+t^{(0,1)}+t^{(-1,2)}+\cdots=t^{(0,1)}+t^{(-1,2)}+\cdots.
\eeq
So $\mathcal T^{(1)}$ goes to zero at $s\to0$. So \Eq{free Schrodinger} may be written as $(\nabla_1^2+\nabla_2^2+\nabla_3^2)\mathcal T^{(1)}=0$ for $p=-1$,
and $\mathcal T^{(1)}$ should satisfy this partial differential equation even at $s_i=0$. Thus $\mathcal T^{(1)}$ must be a harmonic polynomial.
But we do \emph{not} have any nontrivial harmonic polynomial of degree 1 that also satisfies the fermionic antisymmetry.
We are therefore forced to take
\beq
\mathcal{T}^{(1)}=0.
\eeq
So
\beq
t^{(i,j)}=0,~~\text{if}~i+j=1.
\eeq

\paragraph*{\textbf{Step 4.}}
At $s\gg r_e$ we expand $\mathcal{S}^{(0)}$ as
\beq
\mathcal{S}^{(0)}=t^{(0,2)}+t^{(0,1)}+O(s^0)=O(s^0).
\eeq
Combining this with the equation $\widetilde{H}\mathcal{S}^{(0)}=0$, we get
\beq
\mathcal{S}^{(0)}=0.
\eeq
So
\beq
t^{(0,j)}=0.
\eeq

\paragraph*{\textbf{Step 5.}}
At $s\ll R$ we expand $\mathcal{T}^{(0)}$ as
\begin{align}
	\mathcal{T}^{(0)}&=t^{(1,-1)}+t^{(0,0)}+t^{(-1,1)}+t^{(-2,2)}+\cdots\nonumber\\
	&=t^{(1,-1)}+t^{(-1,1)}+\cdots.
\end{align}
$t^{(1,-1)}$ is shown in \Eq{t1,-1}.
$\mathcal{T}^{(0)}$ should satisfy the free Schr\"odinger equation outside of the interaction range, so $(-\nabla_s^2 -3\nabla_R^2/4)\mathcal{T}^{(0)}$ should be equal to some Dirac delta functions that are nonzero at $s_i=0$ only.
$\mathcal{T}^{(0)}$ should also be antisymmetric under the interchange of the fermions.
We have 
$$-\nabla_s ^2 t^{(1,-1)}=8a_p [R_y \partial_x \delta(\vect{s})-R_x \partial_y \delta(\vect{s})],$$
so
\begin{equation}
	\left(-\nabla_\vect{s}^2-\frac{3}{4}\nabla_{\vect{R}}^2 \right)\mathcal{T}_{\vect{s}_1}^{(0)}=8 a_p [R_y \partial_x \delta(\vect{s})-R_x \partial_y \delta(\vect{s})],
\end{equation}
where $\mathcal{T}_{\vect{s}_1}^{(0)}$ is one term of the full $\mathcal{T}^{(0)}$. 
Solving the above equation, we get
\begin{equation}
	\mathcal{T}_{\vect{s}_1}^{(0)}=-\frac{4a_p}{\pi s_1^2}\left(s_x R_y-s_y R_x\right).
\end{equation}
The full $\mathcal{T}^{(0)}$ should also be antisymmetric under the interchange of the fermions, so
\begin{equation}
	\mathcal{T}^{(0)}=-\frac{4a_p}{\pi} \left(s_x R_y-s_y R_x\right) \bigg(\frac{1}{s_1^2}+\frac{1}{s_2^2}+\frac{1}{s_3^2}\bigg).
\end{equation}
If $s\ll R$, we expand $\mathcal{T}^{(0)}$ as $\sum_{n+m=0} t^{(n,m)}$, and get
\begin{subequations}
	\begin{align}
		&t^{(0,0)}=0,\\
		&t^{(-1,1)}=-\frac{8a_p}{\pi R^2}(s_x R_y-s_y R_x),\label{t_-1_1}\\
		&t^{(-2,2)}=0,\\
		&t^{(-3,3)}=\frac{2a_p}{\pi R^6}\left(s_x R_y-s_y R_x\right)(R^2-4 R_s^2)s^2,\label{t_-3_3}\\
		&t^{(-4,4)}=0,\\
		&t^{(-5,5)}=-\frac{a_p}{2\pi R^{10}}\left(s_x R_y-s_y R_x\right)\nonumber\\
		&\quad\quad\quad\quad\times(R^4-12 R^2 R_s^2 +16 R_s^4)s^4,\label{t_-5_5}\\
		&\cdots \nonumber
	\end{align}
\end{subequations}
where $R_s\equiv \vect R\cdot\hat{\vect s}$.

\paragraph*{\textbf{Step 6.}}
At $s\gg r_e$ we expand $\mathcal{S}^{(-1)}$ as
\begin{equation}
	\mathcal{S}^{(-1)}=t^{(-1,1)}+O(s^0).\label{S-1}
\end{equation}
$\mathcal{S}^{(-1)}$ satisfies the equation
\beq
\widetilde{H}\mathcal{S}^{(-1)}=\frac34\nabla_{\vect R}^2\mathcal{S}^{(1)}=0.
\eeq
From the above equations we deduce
\begin{equation}
	\mathcal{S}^{(-1)}=\frac{1}{R}[d_{+} \phi^{(1+)}+d_{-}\phi^{(1-)}].
\end{equation}
$\mathcal{S}^{(-1)}$ contains only the $p$-wave component, in order to be compatible with \Eq{S-1}.
Expanding $\mathcal{S}^{(-1)}$ at $s\gg r_e$, we get
\begin{equation}
	t^{(-1,1)}=\frac{1}{2R}[d_{+}(s_x+\I s_y)+d_{-}(s_x-\I s_y)].
\end{equation}
Comparing this with \Eq{t_-1_1}, we find
\begin{subequations}
	\begin{align}
		&d_{+}=-\frac{8a_p\I}{\pi R} (R_x-\I R_y),\\
		&d_{-}=\frac{8a_p\I}{\pi R} (R_x+\I R_y).
	\end{align}
\end{subequations}
So
\begin{equation}
	\mathcal{S}^{(-1)}=-\frac{8a_p\I}{\pi R} \Big[y^{(1-)}(\hat{\vect{R}})\phi^{(1+)}(\vect{s})-y^{(1+)}(\hat{\vect{R}})\phi^{(1-)}(\vect{s})\Big] .
\end{equation}
Expanding $\mathcal{S}^{(-1)}$ at $s\gg r_e$, we get
\begin{subequations}
	\begin{align}
		&t^{(-1,0)}=0,\\
		&t^{(-1,-1)}=\frac{32a_p^2}{\pi^2 s^2 R^2}(s_xR_y-s_y R_x),\label{t_-1_-1}\\
		&t^{(1,j)}=0,~~j\le-2.
	\end{align}
\end{subequations}

\paragraph*{\textbf{Step 7.}}
At $s\ll R$ we expand $\mathcal{T}^{(-1)}$ as
\begin{equation}
	\mathcal{T}^{(-1)}=t^{(1,-2)}+t^{(0,-1)}+t^{(-1,0)}+O(s^{1})=O(s^{1}).
\end{equation}
The solution to the equation $(\nabla_1^2+\nabla_2^2+\nabla_3^2)\mathcal{T}^{(-1)}=0$ that is compatible with the above expansion
is
\beq
\mathcal{T}^{(-1)}=0.
\eeq

\paragraph*{\textbf{Step 8.}}
At $s\gg r_e$ we expand $\mathcal{S}^{(-2)}$ as
\begin{align}
	\mathcal{S}^{(-2)}&=t^{(-2,4)}+t^{(-2,3)}+t^{(-2,2)}+t^{(-2,1)}+\sum_{j\le0}t^{(-2,j)}\nn\\
	&=O(s^0).\label{S-2expand}
\end{align}
$\mathcal{S}^{(-2)}$ satisfies the equation
\begin{equation}
	\widetilde{H} \mathcal{S}^{(-2)}=\frac34\nabla_\vect R^2\mathcal{S}^{(0)}=0.
\end{equation}
So we get
\begin{equation}\label{S-2decompose}
	\mathcal{S}^{(-2)}=0.
\end{equation}
So
\beq
t^{(-2,j)}=0.
\eeq

\paragraph*{\textbf{Step 9.}}
At $s\ll R$ we expand $\mathcal{T}^{(-2)}$ as
\begin{align}
	\mathcal{T}^{(-2)}&=
	\sum_{j=-3}^0t^{(-2-j,j)}+O(s^{1})\nonumber\\
	&=t^{(-1,-1)}+O(s^{1}).
\end{align}
Solving the equation $(\nabla_1^2+\nabla_2^2+\nabla_3^2)\mathcal{T}^{(-2)}=0$ (for $s_i\ne0$ only) and using the above expansion, we find
\begin{equation}
	\mathcal{T}^{(-2)}=\frac{32a_p^2}{\pi^2}\left(s_x R_y-s_y R_x\right)\sum_{i=1}^3 \frac{1}{B^2 s_i^2}.
\end{equation}
For $s\ll R$, we expand $\mathcal{T}^{(-2)}$ as $\sum_{n+m=-2} t^{(n,m)}$ and get
\begin{subequations}
	\begin{align}
		&t^{(-2,0)}=0,\\
		&t^{(-3,1)}=\frac{40a_p^2}{\pi^2 R^4}(s_x R_y-s_y R_x),\label{t_-3_1}\\
		&t^{(-4,2)}=0,\\
		&t^{(-5,3)}=-\frac{2a_p^2}{\pi^2 R^8}\left(s_x R_y-s_y R_x\right)(23R^2-32 R_s^2)s^2,\label{t_-5_3}\\
		&\cdots \nonumber
	\end{align}
\end{subequations}

\paragraph*{\textbf{Step 10.}}
At $s\gg r_e$ we expand $\mathcal{S}^{(-3)}$ as
\begin{align}
	\mathcal{S}^{(-3)}=t^{(-3,3)}+t^{(-3,1)}+O(s^0).\label{S-3}
\end{align}
Combining this with the equation
\beq
\widetilde{H}\mathcal{S}^{(-3)}=\frac34\nabla_\vect R^2\mathcal{S}^{(-1)}=0,
\eeq
we get
\beq
\mathcal{S}^{(-3)}=\frac{1}{R^3}\Big[\sum_{\pm}e_{\pm} \phi^{(3\pm)}+\sum_{\pm}g_{\pm}\phi^{(1\pm)}\Big].
\eeq
Here $\mathcal S^{(-3)}$ contains the $p$-wave and $f$-wave components, in order to be compatible with \Eq{S-3}.
Expanding $\mathcal S^{(-3)}$ at $s\gg r_e$, we get
\begin{subequations}
	\begin{align}
		t^{(-3,3)}&=\frac{1}{48 R^3}\big[e_{+}(s_x+\I s_y)^3+e_{-}(s_x-\I s_y)^3\big],\\
		t^{(-3,1)}&=\frac{1}{2 R^3}[g_{+}(s_x+\I s_y)+g_{-}(s_x-\I s_y)].
	\end{align}
\end{subequations}
Comparing these equations with Eqs.~\eqref{t_-3_3} and \eqref{t_-3_1}, we find
\begin{subequations}
	\begin{align}
		&e_{+}=-\frac{48a_p\I}{\pi R^3} (R_x-\I R_y)^3,\\
		&e_{-}=\frac{48a_p\I}{\pi R^3} (R_x+\I R_y)^3,\\
		&g_{+}=\frac{40a_p^2\I}{\pi^2 R} (R_x-\I R_y),\\
		&g_{-}=-\frac{40a_p^2\I}{\pi^2 R} (R_x+\I R_y).
	\end{align}
\end{subequations}
So
\begin{align}
	\mathcal{S}^{(-3)}
	=&-\frac{48a_p\I}{\pi R^3} \Big[y^{(3-)}(\hat{\vect{R}})\phi^{(3+)}(\vect{s})-y^{(3+)}(\hat{\vect{R}})\phi^{(3-)}(\vect{s})\Big] \nonumber\\
	&+\frac{40a_p^2\I}{\pi^2 R^3} \Big[y^{(1-)}(\hat{\vect{R}})\phi^{(1+)}(\vect{s})-y^{(1+)}(\hat{\vect{R}})\phi^{(1-)}(\vect{s})\Big].
\end{align}
Expanding $\mathcal{S}^{(-3)}$ at $s\gg r_e$, we get
\begin{subequations}
	\begin{align}
		&t^{(-3,0)}=0,\\
		&t^{(-3,-1)}=-\frac{160a_p^3}{\pi^3 s^2 R^4}(s_xR_y-s_y R_x),\label{t_-3_-1}\\
		&t^{(-3,-2)}=0,\\
		&t^{(-3,-3)}=\frac{768 a_p a_f \I}{\pi^2 R^6 s^6}[(R_x-\I R_y)^3(s_x+\I s_y)^3\nonumber\\
		&\quad\quad\quad\quad-(R_x+\I R_y)^3(s_x-\I s_y)^3],\label{t_-3_-3}\\
		&t^{(-3,j)}=0,~~j\le-4.
	\end{align}
\end{subequations}

\paragraph*{\textbf{Step 11.}}
At $s\ll R$ we expand $\mathcal{T}^{(-3)}$ as
\beq
\mathcal{T}^{(-3)}=\sum_{j=-4}^0t^{(-3-j,j)}+O(s^1)=O(s^{1}).
\eeq
Solving the equation $(\nabla_1^2+\nabla_2^2+\nabla_3^2)\mathcal{T}^{(-3)}=0$ (for $s_i\ne0$ only) and using the above expansion, we find
\beq
\mathcal{T}^{(-3)}=0.
\eeq
So
\beq
t^{(i,j)}=0,~~\text{if}~i+j=-3.
\eeq

\paragraph*{\textbf{Step 12.}}
At $s\gg r_e$ we expand $\mathcal{S}^{(-4)}$ as
\begin{align}
	\mathcal{S}^{(-4)}&=t^{(-4,6)}+t^{(-4,5)}+...+t^{(-4,1)}+O(s^0)\nn\\
	&=O(s^0).\label{S-4expand}
\end{align}
$\mathcal{S}^{(-4)}$ satisfies the equation
\begin{equation}
	\widetilde{H} \mathcal{S}^{(-4)}=\frac34\nabla_\vect R^2\mathcal{S}^{(-2)}=0.
\end{equation}
So we get
\begin{equation}\label{S-4decompose}
	\mathcal{S}^{(-4)}=0.
\end{equation}
So
\beq
t^{(-4,j)}=0.
\eeq

\paragraph*{\textbf{Step 13.}}
At $s\ll R$ we expand $\mathcal{T}^{(-4)}$ as
\begin{align}
	\mathcal{T}^{(-4)}&=\sum_{j=-5}^0t^{(-4-j,j)}+O(s^1)\nn\\
	&=-\frac{160a_p^3}{\pi^3 s^2 R^4}(s_xR_y-s_y R_x)+O(s^{1}).
\end{align}
Solving the equation $(\nabla_1^2+\nabla_2^2+\nabla_3^2)\mathcal{T}^{(-4)}=0$ (for $s_i\ne0$ only) and using
the above expansion, we find
\begin{align}
	\mathcal{T}^{(-4)}=&(s_x R_y-s_y R_x)\Bigg[-\frac{3D_F}{2\pi^2 B^6}\nn\\
	&-\frac{160a_p^3}{\pi^3 B^6}\sum_i\bigg(\frac{R_i^2}{s_i^2}+3\ln\frac{B^2}{\sqrt{|a_p|}s_i}\bigg)\Bigg], 
\end{align}
where $D_F$ is a new coefficient and we call it the three-body scattering hypervolume. Its value depends on the details of the interactions.

Expanding $\mathcal{T}^{(-4)}$ at $s\ll R$ as $\sum_{i+j=-4} t^{(i,j)}$, we get
\begin{subequations}
	\begin{align}
		&t^{(-4,0)}=0,\\
		&t^{(-5,1)}=(s_x R_y-s_y R_x) \bigg(-\frac{3D_F}{2\pi^2 R^6}\nonumber\\
		&\quad\quad+\frac{280a_p^3}{\pi^3 R^6}-\frac{480a_p^3}{\pi^3 R^6}\ln \frac{R^4}{|a_p|^{3/2}s}\bigg),\label{t_-5_1}\\
		&\cdots \nonumber
	\end{align}
\end{subequations}

\paragraph*{\textbf{Step 14.}}
At $s\gg r_e$ we expand $\mathcal{S}^{(-5)}$ as
\begin{align}
	\mathcal{S}^{(-5)}&=\sum_{j\le 7}t^{(-5,j)}\nn\\
	&=t^{(-5,5)}+t^{(-5,3)}+t^{(-5,1)}+O(s^0).
\end{align}
Combining this with the equation
\begin{align}
	&\widetilde{H}\mathcal{S}^{(-5)}=\frac34\nabla_\vect R^2\mathcal{S}^{(-3)}\nn\\
	&=\frac{240 a_p^2 \I}{\pi^2 R^6}[(R_x-\I R_y)\phi^{(1+)}-(R_x+\I R_y)\phi^{(1-)}],
\end{align}
we get
\begin{equation}
	\begin{split}
		\mathcal{S}^{(-5)}&=\frac{240 a_p^2 \I}{\pi^2 R^5}[y^{(1-)}(\widehat{R})f^{(1+)}(s)-y^{(1+)}(\widehat{R})f^{(1-)}(s)]\\
		&-\frac{960a_p\I}{\pi R^5} [y^{(5-)}(\widehat{R})\phi^{(5+)}(s)-y^{(5+)}(\widehat{R})\phi^{(5-)}(s)]\\
		&+\frac{384a_p^2 \I}{\pi^2 R^5}[y^{(3-)}(\widehat{R})\phi^{(3+)}(s)-y^{(3+)}(\widehat{R})\phi^{(3-)}(s)]\\
		&-\frac{\zeta\I}{R^5}[y^{(1-)}(\widehat{R})\phi^{(1+)}(s)-y^{(1+)}(\widehat{R})\phi^{(1-)}(s)],
	\end{split}
\end{equation}
where
\begin{equation}
	\zeta=\frac{3D_F}{2\pi^2}-\frac{280 a_p^3}{\pi^3}+\frac{1920a_p^3}{\pi^3}\ln\frac{R}{\widetilde{R}}.
\end{equation}
We have thus derived the 111 expansion to the order $B^{-4}$ and the 21 expansion to the order $R^{-5}$.

\section{The Born expansion of the three-body wave function\label{appendix:Born}}
For weak interaction potentials, we can expand the three-body wave function as a Born series:
\begin{equation}
	\Psi=\Psi_0+\widehat{G}\mathcal{V}\Psi_0+(\widehat{G}\mathcal{V})^2\Psi_0+\cdots,
\end{equation}
where $\Psi_0=s_x R_y-s_y R_x$ is the wave function of three free fermions, $\mathcal{V}$ is the interaction potential, and $\widehat{G}=-\widehat{H}_0^{-1}$ is the Green's operator, $\widehat{H}_0$ is the three-body kinetic energy operator.
We define $\mathcal{V}=V_3(s_1,s_2,s_3)+\sum_{i=1}^3 V(s_i)$.
We assume that $V(s)$ vanishes at $s>r_e$ and that $V_3(s_1,s_2,s_3)$ vanishes if $s_1>r_e$ or $s_2>r_e$ or $s_3>r_e$.
\subsection{The first-order term}
The first-order term in the Born series is
\begin{align}
	\Psi_1(\bm{\xi})&=\widehat{G}\mathcal{V}\Psi_0=\widehat{G}V_3 \Psi_0+\sum_{i=1}^3\widehat{G}V^{(i)} \Psi_0\nonumber\\
	&=\frac{m}{\hbar^2}\int\!\!d^4\xi'\,\mathcal{G}(\bm{\xi}-\bm{\xi}')V_3(s_1',s_2',s_3')\Psi_0(\bm{\xi}')\nonumber\\
	&\quad+\frac{m}{\hbar^2}\sum_{i=1}^3 \int \!\!d^4 \xi' ~\mathcal{G}(\bm{\xi}-\bm{\xi}')~V^{(i)}(\bm{\xi}') \Psi_0(\bm{\xi}'),
\end{align}
where $\bm\xi=(\vect{s},2\vect{R}/\sqrt{3})$ and $\bm{\xi}'=(\vect{s}',2\vect{R}'/\sqrt{3})$ are 4 dimensional vectors, $\mathcal{G}$ is the Green's function in 4-dimensional space,
\begin{equation}
	\mathcal{G}(\bm{\xi}-\bm{\xi}')=-\frac{1}{4\pi^2|\bm{\xi}-\bm{\xi}'|^2},
\end{equation}
and $V^{(i)}(\bm\xi')=V(s'_i)$.
$\vect{s}_1'=\vect{s}'$, $\vect{s}_2'=-\frac{1}{2}\vect{s}'-\vect{R}'$, $\vect{s}_3'=-\frac{1}{2}\vect{s}'+\vect{R}'$.
We write $\Psi_1$ as $\widehat{G}V_3\Psi_0+\sum_{i=1}^3\Psi_{1}^{(i)}$, where
\begin{align}
	\Psi_{1}^{(i)}&\equiv\widehat{G}V^{(i)}\Psi_0=\int \!\!d^4 \xi' ~\frac{-m}{4\pi^2\hbar^2|\bm{\xi}-\bm{\xi}'|^2}V^{(i)}(\bm{\xi}') ~\Psi_0(\bm{\xi}')\nonumber\\
	&=\frac{4}{3}\int\!\! d^2 s' \!\!\int\!\! d^2 R' \frac{-mV(s')(s_x' R_y'-s_y' R_x')}{4\pi^2\hbar^2[(\vect{s}_i-\vect{s}')^2+\frac{4}{3}(\vect{R}_i-\vect{R}')^2]}.
\end{align}
Finishing this integral and taking the sum over $i$, we get
\begin{equation}\label{1-order}
	\sum_{i=1}^3\Psi_1^{(i)}=-\frac{1}{2}(s_x R_y-s_y R_x)\sum_{i=1}^3\left[\frac{\alpha_2 (s_i)}{s_i^2}+\bar{\alpha}_0(s_i)\right],
\end{equation}
where
\begin{subequations}
	\begin{align}
		&\alpha_n(s)      \equiv \frac{m}{\hbar^2}\int_{0}^{s}     \!\!\! ds'~ s'^{n+1}V(s'),\\
		&\bar{\alpha}_n(s)\equiv \frac{m}{\hbar^2}\int_{s}^{\infty}\!\!\! ds'~ s'^{n+1}V(s').
	\end{align}
\end{subequations}
Since $V(s)$ is a finite-range potential which vanishes at $s>r_e$, we have $\alpha_n(s)=\alpha_n\equiv m\int_0^{\infty}ds' s'^{n+1}V(s')/\hbar^2$ and $\bar{\alpha}_n(s)=0$ if $s>r_e$.
If all three $s_i>r_e$, $\sum_{i=1}^3\Psi_1^{(i)}$ is simplified as
\begin{equation}
	\sum_{i=1}^3\Psi_1^{(i)}=-(s_x R_y-s_y R_x)\sum_{i=1}^3\frac{\alpha_2}{2s_i^2}.
\end{equation}

\begin{align}\label{GV3}
&\widehat{G}V_3\Psi_0=\int \!\!d^4 \xi' ~\frac{-m V_3(\bm{\xi}') ~\Psi_0(\bm{\xi}')}{4\pi^2\hbar^2|\bm{\xi}-\bm{\xi}'|^2}\nonumber\\
	&=\frac{4}{3}\int\!\! d^2 s' \!\!\int\!\! d^2 R' \frac{-mV_3(s_1',s_2',s_3')(s_x' R_y'-s_y' R_x')}{4\pi^2\hbar^2[(\vect{s}-\vect{s}')^2+\frac{4}{3}(\vect{R}-\vect{R}')^2]}.
\end{align}
Let $s_x'=s'\cos \alpha'$, $s_y'=s'\sin \alpha'$, $R_x'=R'\cos(\alpha'+\theta')$, $R_y'=R'\sin(\alpha'+\theta')$, $s_x=s \cos\alpha$, $s_y=s\sin\alpha$, $R_x=R\cos(\alpha+\theta)$, $R_y=R\sin(\alpha+\theta)$. We have
\begin{equation}
\int\!\! d^2 s' \!\!\int\!\! d^2 R'=\int_0^{\infty}\!\! ds' s' \int_0^{\infty}\!\! dR' R' \int_{-\pi}^{\pi}\!\! d\theta' \int_{-\pi}^{\pi}\!\! d\alpha'.
\end{equation}
Since $V_3$ is a finite-range potential, the integral on the right hand side of \Eq{GV3} may be expanded when $s$ and $R$ go to infinity for any fixed ratio $s/R$.
Expanding this integral at large $B$, and using the fact that $V_3$ is an even function of $\theta'$ and is independent of $\alpha'$, we get
%
\begin{widetext}
\begin{align}
    \widehat{G}V_3\Psi_0&= -\frac{3mRs\sin\theta}{\pi \hbar^2 B^6}\int_{0}^{\infty}\!\!ds'\int_{0}^{\infty}\!\! dR'\int_{0}^{\pi}\!\! d\theta'~s'^3 R'^3 \sin^2\theta'~V_3(s_1',s_2',s_3')+O(B^{-6})  \nonumber\\
    &=-\frac{6m(s_x R_y-s_y R_x)}{\pi \hbar^2 B^6}\int ds_1' ds_2' ds_3'~s_1's_2's_3' V_3 (s_1',s_2',s_3')S_{\Delta} (s_1',s_2',s_3')+O(B^{-6}).\label{GV3Psi0}
\end{align}
In the first line of \Eq{GV3Psi0},
$s_1'=s'$, $s_2'=\sqrt{R'^2+\frac{1}{4}s'^2+R's'\cos\theta'}$, and $s_3'=\sqrt{R'^2+\frac{1}{4}s'^2-R's'\cos\theta'}$.

\subsection{The second-order term}
The second-order term in the Born series is
\begin{equation}
	\Psi_2=\widehat{G}\mathcal{V}\Psi_1=\sum_{ij}\widehat{G}V^{(i)} \widehat{G}V^{(j)}\Psi_0+\sum_i \widehat{G}V^{(i)}\widehat{G}V_3\Psi_0+\sum_i \widehat{G}V_3\widehat{G}V^{(i)}\Psi_0+(\widehat{G}V_3)^2\Psi_0.\label{2nd-order}
\end{equation}
We define
\begin{equation}
	\Psi_{2}^{(ij)}\equiv \widehat{G}V^{(i)} \widehat{G}V^{(j)}\Psi_0=\int \!\!d^4 \xi' ~\frac{-m}{4\pi^2\hbar^2|\bm{\xi}-\bm{\xi}'|^2}V^{(i)}(\bm\xi')\Psi_{1}^{(j)}(\bm{\xi}').
\end{equation}
In particular,

	\begin{equation}
		\Psi_{2}^{(ii)}=\frac{4}{3}\int\!\! d^2 s' \!\!\int\!\! d^2 R' \frac{-mV(s')(s_x' R_y'-s_y' R_x')}{4\pi^2\hbar^2[(\vect{s}_i-\vect{s}')^2+\frac{4}{3}(\vect{R}_i-\vect{R}')^2]}\left(-\frac{1}{2}\right)\left[\frac{\alpha_2 (s')}{s'^2}+\bar{\alpha}_0(s')\right].
	\end{equation}
	If all three $s_i>r_e$, we can evaluate the integral to obtain
	\begin{equation}
		\Psi_{2}^{(ii)}=(s_x R_y-s_y R_x)\frac{\beta_2}{2s_i^2},\label{psi2-ii}
	\end{equation}
	where $\beta_2$ is defined as
	\begin{equation}
		\beta_2\equiv \frac{m^2}{\hbar^4}\int_{0}^{\infty}\!\!\! ds\int_{0}^{s}\!\!\! ds' ~s s'^3 V(s) V(s').
	\end{equation}
	
	If $j\neq i$, we have two different values of $j$ for each $i$. For a given value of $i$,
	\begin{align}
		\sum_{j\neq i}\Psi_{2}^{(ij)}&=\frac{4}{3}\int\!\! d^2 s' \!\!\int\!\! d^2 R' 
		\frac{mV(s')(s_x' R_y'-s_y' R_x')}{8\pi^2\hbar^2[(\vect{s}_i-\vect{s}')^2+\frac{4}{3}(\vect{R}_i-\vect{R}')^2]} \nonumber\\
		&\quad\times\left[\frac{\alpha_2 (|\vect{R}'-\frac{1}{2}\vect{s}'|)}{|\vect{R}'-\frac{1}{2}\vect{s}'|^2}+\bar{\alpha}_0(|\vect{R}'-\frac{1}{2}\vect{s}'|)+\frac{\alpha_2 (|\vect{R}'+\frac{1}{2}\vect{s}'|)}{|\vect{R}'+\frac{1}{2}\vect{s}'|^2}+\bar{\alpha}_0(|\vect{R}'+\frac{1}{2}\vect{s}'|)\right]\nonumber\\
		&=\int\!\! d^2 s' \!\!\int\!\! d^2 R' \frac{mV(s')}{8\pi^2\hbar^2}(s_x'R_y'-s_y' R_x')\left[\frac{\alpha_2 (R')}{R'^2}+\bar{\alpha}_0(R')\right]\nonumber\\
		&\quad\times\left[\frac{1}{\frac{3}{4}(\vect{s}_i-\vect{s}')^2+(\vect{R}_i-\vect{R}'-\frac{1}{2}\vect{s}')^2}+\frac{1}{\frac{3}{4}(\vect{s}_i-\vect{s}')^2+(\vect{R}_i-\vect{R}'+\frac{1}{2}\vect{s}')^2}\right].
	\end{align}
	We split this integral into two parts by writing
	\begin{equation}
		\frac{\alpha_2 (R')}{R'^2}+\bar{\alpha}_0(R')=\frac{\alpha_2}{R'^2}+\left[ \frac{\alpha_2(R')-\alpha_2}{R'^2}+\bar{\alpha_0}(R')\right] 
		=f_L(R')+f_S(R'),
	\end{equation}
	where $f_S$ is a short-range function and $f_L$ is a long-range function of $R'$:
	\begin{subequations}
		\begin{align}
			&f_S(R')=\frac{\alpha_2(R')-\alpha_2}{R'^2}+\bar{\alpha}_0(R')=\bar{\alpha}_0(R')-\frac{\bar{\alpha}_2(R')}{R'^2},\\
			&f_L(R')=\frac{\alpha_2}{R'^2}.
		\end{align}
	\end{subequations}
	The integral containing $f_S$ is
	\begin{align}
		&\int\!\! d^2 s' \!\!\int\!\! d^2 R' \frac{mV(s')}{8\pi^2\hbar^2}(s_x'R_y'-s_y' R_x')f_S(R')\left[\frac{1}{\frac{3}{4}(\vect{s}_i-\vect{s}')^2+(\vect{R}_i-\vect{R}'-\frac{1}{2}\vect{s}')^2}+\frac{1}{\frac{3}{4}(\vect{s}_i-\vect{s}')^2+(\vect{R}_i-\vect{R}'+\frac{1}{2}\vect{s}')^2}\right]\nonumber\\
		&=\int\!\! d^2 s' \!\!\int\!\! d^2 R' \frac{mV(s')}{8\pi^2\hbar^2}(s_x'R_y'-s_y' R_x')\left[\bar{\alpha}_0(R')-\frac{\bar{\alpha}_2(R')}{R'^2}\right] \left[ \frac{12(\vect{s}_i\cdot\vect{s}')(\vect{R}_i\cdot\vect{R}')}{(R_i^2+\frac{3}{4}s_i^2)^3}+...\right] \nonumber\\
		&=-\frac{3\alpha_2\alpha_4}{8B^6}(s_x R_y-s_y R_x)+O(B^{-6}),\label{f-S}
	\end{align}
	where we have expanded the integral to the leading-order term assuming that $s$ and $R$ go to infinity simultaneously.
	The integral containing $f_L$ is
	\begin{align}
		&\int\!\! d^2 s' \!\!\int\!\! d^2 R' \frac{mV(s')}{8\pi^2\hbar^2}(s_x'R_y'-s_y' R_x')f_L(R')\left[\frac{1}{\frac{3}{4}(\vect{s}_i-\vect{s}')^2+(\vect{R}_i-\vect{R}'-\frac{1}{2}\vect{s}')^2}+\frac{1}{\frac{3}{4}(\vect{s}_i-\vect{s}')^2+(\vect{R}_i-\vect{R}'+\frac{1}{2}\vect{s}')^2}\right]\nonumber\\
		&=(s_x R_y-s_y R_x)\left(\frac{\alpha_2^2}{2B^2s_i^2}+\frac{3\alpha_2\alpha_4}{16 B^6}\right) +O(B^{-6}).\label{f-L}
	\end{align}
	
	Combining the above results, we get
	\begin{equation}
		\sum_{j\neq i}\Psi_{2}^{(ij)}=(s_x R_y-s_y R_x)\left(\frac{\alpha_2^2}{2B^2s_i^2}-\frac{3\alpha_2\alpha_4}{16 B^6}\right) +O(B^{-6})\label{psi2-ij}
	\end{equation}
	for any given value of $i$.
	Substituting Eqs.~\eqref{psi2-ii} and \eqref{psi2-ij} into \Eq{2nd-order}, we get
	\begin{equation}
		\Psi_2=(s_x R_y-s_y R_x)\cdot\bigg\{\Big[\sum_{i=1}^3 \left( \frac{\beta_2}{2 s_i^2}+\frac{\alpha_2^2}{2B^2 s_i^2}\right)\Big] -\frac{9\alpha_2\alpha_4}{16 B^6}\bigg\}+O(B^{-6})
		+O(VV_3)+O(V_3^2).
	\end{equation}
We have not evaluated the terms $\sum_i \widehat{G}V^{(i)}\widehat{G}V_3\Psi_0+\sum_i \widehat{G}V_3\widehat{G}V^{(i)}\Psi_0$ and
$(\widehat{G}V_3)^2\Psi_0$, which are of order $VV_3$ and $V_3^2$, respectively.
	
\end{widetext}

\bibliography{ref}

\end{document}